\documentclass[reprint,amsmath,amssymb,aps,]{revtex4-1}
\usepackage{xcolor}
\usepackage{siunitx}
\usepackage{graphicx}
\usepackage{dcolumn}
\usepackage{bm}

\begin{document}

\preprint{APS/123-QED}

\title{Tuning the structural and antiferromagnetic phase transitions in UCr$_{2}$Si$_2$: \\hydrostatic pressure and chemical substitution}

\author{Y. Lai$^{1,2}$, K. Wei$^{1}$, G. Chappell$^{1,2}$, J. Diaz$^{3}$, T. Siegrist$^{1,2}$, P. J. W. Moll$^{3}$, D. Graf$^{1}$, R. E. Baumbach$^{1,2}$}

\email{baumbach@magnet.fsu.edu}
\email{Y. Lai works at LANL now}
\affiliation{$^1$National High Magnetic Field Laboratory, Florida State University, Tallahassee, FL 32310, USA}
\affiliation{$^2$Department of Physics, Florida State University, Tallahassee, FL 32306, USA}
\affiliation{$^3$Institute of Materials (IMX), EPFL, Lausanne, Switzerland}

\date{\today}
\begin{abstract}
Structural phase transitions in $f$-electron materials have attracted sustained attention both for practical and basic science reasons, including that they offer an environment to directly investigate relationships between structure and the $f$-state. Here we present results for UCr$_2$Si$_2$, where structural (tetragonal $\rightarrow$ monoclinic) and antiferromagnetic phase transitions are seen at $T_{\rm{S}}$ $=$ 205 K and $T_{\rm{N}}$ $=$ 25 K, respectively. We also provide evidence for an additional second order phase transition at $T_{\rm{X}}$ = 280 K.  We show that $T_{\rm{X}}$, $T_{\rm{S}}$, and $T_{\rm{N}}$ respond in distinct ways to the application of hydrostatic pressure and Cr $\rightarrow$ Ru chemical substitution. In particular, hydrostatic compression increases the structural ordering temperature, eventually causes it to merge with $T_{\rm{X}}$ and destroys the antiferromagnetism. In contrast, chemical substitution in the series UCr$_{2-x}$Ru$_x$Si$_2$ suppresses both $T_{\rm{S}}$ and $T_{\rm{N}}$, causing them to approach zero temperature near $x$ $\approx$ 0.16 and 0.08, respectively. The distinct $T-P$ and $T-x$ phase diagrams are related to the evolution of the rigid Cr-Si and Si-Si substructures, where applied pressure semi-uniformly compresses the unit cell and Cr $\rightarrow$ Ru substitution results in uniaxial lattice compression along the tetragonal $c$-axis and an expansion in the $ab$-plane. These results provide insights into an interesting class of strongly correlated quantum materials where degrees of freedom associated with $f$-electron magnetism, strong electronic correlations, and structural instabilities are readily controlled.  

\begin{description}

\item[PACS numbers]
PACS

\end{description}
\end{abstract}

\pacs{Valid PACS appear here}

\maketitle

\section{\label{sec:level1}Introduction}
The $f$-electron state is a main driver of the chemistry and physics in actinide-based intermetallics, and influences properties ranging from the functional (e.g., crystal structure/density, melting temperature, and thermal conductivity) to the exotic (e.g., unconventional superconductivity, magnetism, and electronic topology)~\cite{moore,Pfleiderer09,Maple10,dzero}. This is due to several factors, including (i) the remarkable flexibility of the $f$-electron valence, which readily conforms to a multitude of different crystal/chemical environments and (ii) that $f$-electrons intrinsically carry unique features including a tendency to hybridize with conduction electrons and strong spin-orbit coupling that can produce large spin anisotropy. An intriguing subset of behaviors that such materials exhibit are structural/electronic instabilities: e.g., as for cerium, which undergoes an isostructural volume collapse between two face centered cubic structures under applied pressure~\cite{zhao_1997} and plutonium, which features the largest number of distinct structural phases in the temperature-pressure phase diagram of all elements~\cite{moore,Wick_1967}. In the  broader family of actinide based materials, these features have implications for development of nuclear fuels~\cite{Gofryk_2017}, waste forms for storage of dangerous isotopes~\cite{Tom_2015}, and other applications ~\cite{application}. 

The system UCr$_2$Si$_2$ offers the chance for an improved understanding of the relationship between structure and the $f$-state, as it was earlier shown to crystallize in the ubiquitous ThCr$_2$Si$_2$ tetragonal structure and to undergo a phase transition to a monoclinic phase near $T_{\rm{S}}$ $\approx$ 205 K~\cite{Matsuda1_2002,Matsuda1_2003,Lemoine_2018}. The U ions carry a magnetic moment, the magnetic anisotropy is Ising-like, there is evidence for Kondo lattice behavior, and antiferromagnetic ordering appears at $T_{\rm{N}}$ $\approx$ 25 K. The origin of the structural phase transition has not been established, but it is noteworthy that (i) in the broader family ($Ln/An$)Cr$_2$Si$_2$ ($Ln$ $=$ lanthanide and $An$ $=$ actinide) the Cr ion carries a magnetic moment (here it does not) and (ii) no other example exhibits a structural phase transition~\cite{Klosek_2008, Ryan_2003, Shatruk_2019,NpCr2Si2_1977,Book_1998}. This might suggest that this compound is differentiated from its analogues mainly due to strong hybridization between the uranium $f$-electrons and the conduction electron, which then produces the structural instability. On the other hand, it was earlier proposed that a phonon driven instability associated with the Cr-Si and Si-Si bonding is responsible for this behavior~\cite{Lemoine_2018}. In addition to understanding the origin for the structural phase transition, it is attractive to consider the possibility of driving it from the thermally driven (classical) regime to the zero temperature (quantum) regime, as has been done earlier for some magnetic correlated electron materials that conform to the semi-universal quantum critical point phase diagram~\cite{Rosch07,Gegenwart_08,brando16,Lee_2006,Stewart_2011,Singleton_2002}.

In order to clarify the behavior of this compound and to assess whether a structural quantum critical point might be induced, we have carried out measurements of polycrystalline UCr$_2$Si$_2$ under applied pressure and also for specimens where Cr is substituted by Ru. These measurements are complementary; while applied pressure is expected to compress the unit cell volume, X-Ray diffraction measurements show that Cr $\rightarrow$ Ru substitution results in compression along the $c$-axis and expansion in the $ab$-plane. In the former case, the structural phase transition is initially unchanged, but starting near $P$ $=$ 7.6 kbar it undergoes a rapid evolution, increases in temperature, and eventually merges with a previously overlooked higher temperature feature at $T_{\rm{X}}$ = 280 K. In contrast, Cr $\rightarrow$ Ru substitution quickly obscures $T_{\rm{X}}$ and drives a smooth suppression of $T_{\rm{S}}$ until it rapidly collapses, possibly towards a structural quantum phase transition near $x_{\rm{c,S}}$ $\approx$ 0.16. The antiferromagnetism also evolves in a complex way; under hydrostatic pressure it abruptly disappears as $T_{\rm{S}}$ begins to increase, while it is gradually suppressed with increasing $x$ and collapses at a value $x$ $<$ $x_{\rm{c,S}}$. Together, these features provide a setting in which to investigate the interplay between magnetism, structure, and Kondo lattice behavior and thereby sets the stage for developing a previously unexplored class of strongly correlated quantum materials.

\section{\label{sec:level1}Experimental Methods}
Polycrystalline specimens of UCr$_{2-x}$Ru$_x$Si$_2$ were grown from elements with purities \(>\) 99.9\% by combining elements in the molar ratio U:Cr:Ru:Si $;$ 1:(2-$x$):$x$:2 using an arc-furnace. To facilitate mixing of the elements, the U and Si were melted first, then the resulting mixture was combined with Cr and Ru. The crystal structure and chemical composition were verified by powder X-Ray diffraction (XRD) and energy dispersive spectroscopy (EDS) analysis. EDS results are shown in Fig. S1, where the measured concentration $x_{\rm{act}}$ is compared to the nominal concentration $x_{\rm{nom}}$. Throughout the rest of the manuscript we use $x_{\rm{act}}$ unless otherwise specified. Powder XRD measurements were performed at temperatures 300 K, 40 K, and 10 K using a Guinier diffractometer. Ambient pressure electrical resistance $R$ measurements for temperatures $T$ $=$ 0.5 $-$ 300 K were performed in a four-wire configuration and the heat capacity $C$ was measured for $T$ $=$ 1.8 $-$ 250 K using a Quantum Design Physical Property Measurement System. $R(T)$ measurements under applied pressure were performed using a piston cylinder pressure cell with Daphne 7474 oil as the pressure transmitting medium. The pressure was determined by the shift in ruby flourescence peaks as measured below $T$ $=$ 10 K. 

\section{\label{sec:level1}Results}
\begin{figure}[h]
	\begin{center}
		\includegraphics[width=1\linewidth]{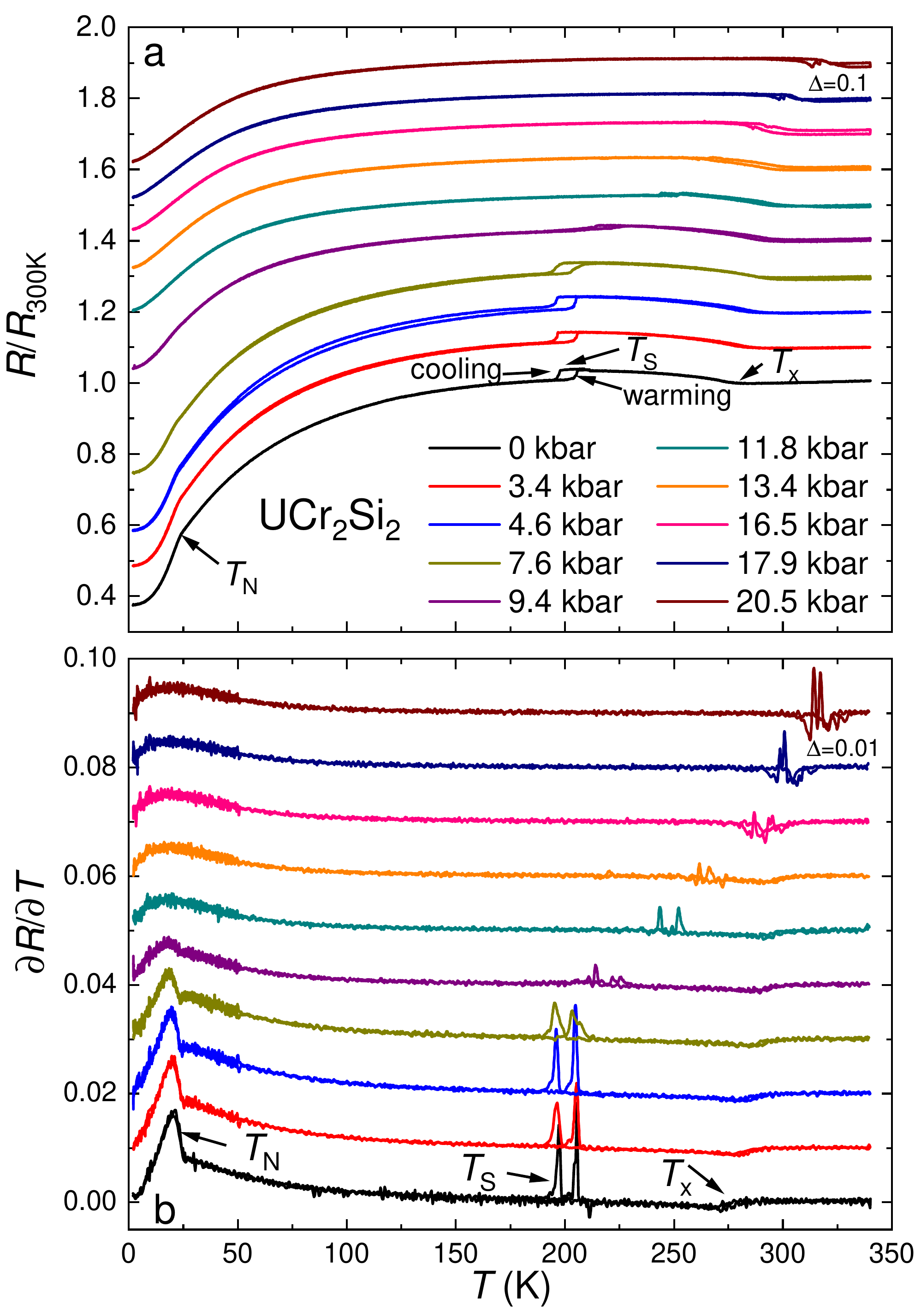}
		\caption{(a) Electrical resistance divided by the room temperature value $R/R_{300K}$ vs temperature $T$ for UCr$_2$Si$_2$ under applied pressure. The anomaly at $T_{\rm{X}}$, the structural phase transition $T_{\rm{S}}$, and the antiferromagnetic phase transition $T_{\rm{N}}$ are indicated by arrows. For clarity, the curves are vertically offset by the value $\Delta$ $=$ 0.1.
		(b) The derivative of $R/R_{300K}$ with respect to $T$ $\partial$$R$/$\partial$$T$ vs $T$ for UCr$_2$Si$_2$ under applied pressure. Curves are offset by constant values.}
		\label{pressure}
	\end{center}
\end{figure}

Electrical resistance measurements $R/R_{300K}$ under hydrostatic pressure were performed in order to establish the influence of isotropic volume compression (Fig.~\ref{pressure}). For $P$ $=$ 0, $R/R_{300K}$ vs. $T$ is consistent with earlier reports ~\cite{Matsuda1_2002,Matsuda1_2003,Lemoine_2018}, where the structural phase transition at $T_{\rm{S}}$ $=$ 205 K and the antiferromagnetic phase transition at $T_{\rm{N}}$ $=$ 25 K appear as a hysteretic decrease and a knee-like reduction in $R/R_{300K}$, respectively. These features are superimposed on a broader Kondo lattice-like temperature dependence. We additionally observe an abrupt change in slope near $T_{\rm{X}}$ $=$ 280 K which was earlier seen by Matsuda $et$ $al$, but was not discussed.~\cite{Matsuda1_2002,Matsuda1_2003} As shown in Figs. S2 and S3; (i) it is present in $\rho(T)$ measurements for both polycrystal and single crystal specimens, (ii) it is anisotropic in $\rho(T)$, and (iii) it appears as a knee-like feature in the heat capacity, which indicates that it is a bulk thermodynamic phase. We also note that it has a close similarity to the charge density wave features seen in $R$Pt$_2$Si$_2$($R$ = Y, La, Nd, Lu)~\cite{LaPt2Si2, RPt2Si2} and UPt$_2$Si$_2$~\cite{UPt2Si2}. 

Pressure initially has a limited effect on $T_{\rm{X}}$, $T_{\rm{S}}$, and $T_{\rm{N}}$ but starting above 7.6 kbar they each undergo a clear evolution. In particular, $T_{\rm{S}}$ rapidly moves to higher temperatures, loses magnitude, and finally merges with $T_{\rm{X}}$. This suggests that compression of the unit cell volume has a tendency to stabilize both of the high temperature phase transitions, which might be tied to an increasing lattice stiffness. However, above 7.6 kbar the feature at $T_{\rm{S}}$ changes shape and eventually becomes difficult to distinguish from $T_{\rm{X}}$ as they merge together. This may suggest a transformation into another structural phase over this $P$ range. At lower temperatures $T_{\rm{N}}$ is gradually suppressed and abruptly disappears above 9.4 kbar, showing that the magnetic order is tied directly to the monoclinic structure, which may support the view that the structure changes at high $P$. Finally, we note that the Kondo lattice behavior persists at all $P$. 

\begin{figure}[h]
	\begin{center}
		\includegraphics[width=1\linewidth]{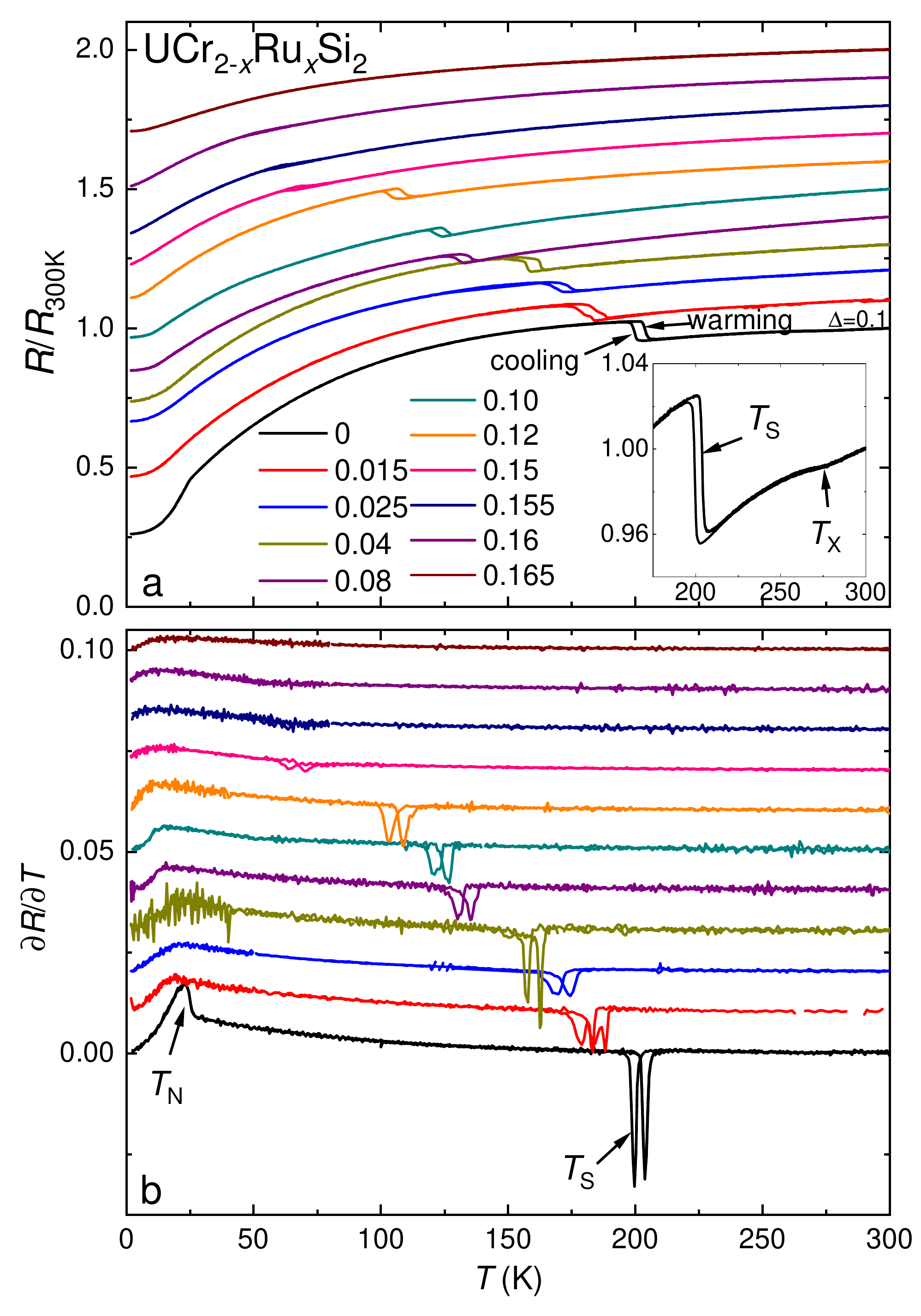}
		\caption{(a) Electrical resistance divided by the room temperature value $R/R_{300K}$ vs temperature $T$ for UCr$_{2-x}$Ru$_x$Si$_2$. For clarity, the curves are vertically offset by the value $\Delta$ $=$ 0.1. (inset) Zoom of $R(T)$ for $x$ $=$ 0 showing the features at $T_{\rm{X}}$ and $T_{\rm{S}}$.
		(b) The derivative of $R/R_{300K}$ with respect to $T$ $\partial$$R$/$\partial$$T$ vs $T$ for UCr$_{2-x}$Ru$_x$Si$_2$. Curves are offset by constant value $\Delta$ $=$ 0.01. The anomalies $T_{\rm{S}}$ and  $T_{\rm{N}}$ are labeled.
		}
		\label{substitution}
	\end{center}
\end{figure}

The influence of non-isoelectronic chemical substitution (Cr $\rightarrow$ Ru) on UCr$_2$Si$_2$ is revealed in the $R(T)$ curves shown in Fig.~\ref{substitution}. For $x$ $=$ 0, the feature at $T_{\rm{S}}$ results in an increase in $R/R_{300K}$ and the change at $T_{\rm{X}}$ is weak (Fig.~\ref{substitution}a inset). This contrasts with results for the $x$ $=$ 0 specimen that was used for measurements under $P$ (Fig.~\ref{pressure}), but is consistent with what is seen for single crystal specimens (Fig. S3) where $R/R_{300K}$ is anisotropic depending on whether the electrical current is applied in the $ab$-plane or along the $c$-axis. In particular, $R(T)$ increases (or decreases) at $T_{\rm{S}}$ for electrical current applied in the $ab$ plane (or along the $c$-axis) and $T_{\rm{X}}$ is only easily observed when the electrical current is applied along the $c$-axis.~\cite{Matsuda1_2002,Matsuda1_2003} 

For the chemical substitution series, we find that the feature at $T_{\rm{X}}$ rapidly disappears and the one at $T_{\rm{S}}$ is smoothly suppressed up to $x$ $=$ 0.16, where it may collapse towards zero temperature. The disappearance of $T_{\rm{X}}$ might indicate that this phase transition is unstable against chemical substitution, but it is also possible that it merely becomes unobservable due to broadening. As $T_{\rm{S}}$ is suppressed it retains its characteristic shape and hysteresis, showing that it remains first order throughout the phase diagram. We also note that the feature at $T_{\rm{N}}$ is suppressed with increasing $x$ and is no longer observable near $x$ $=$ 0.08. The heat capacity and magnetic susceptibility data presented below also reveal a systematic suppression of the magnetic ordering with chemical substitution, resulting in its disappearance for $x$ $\approx$ 0.08 - 0.10. Across the entire substitution series, the underlying Kondo lattice behavior is preserved. 

\begin{figure*}[!tht]
	\begin{center}
		\includegraphics[width=1\linewidth]{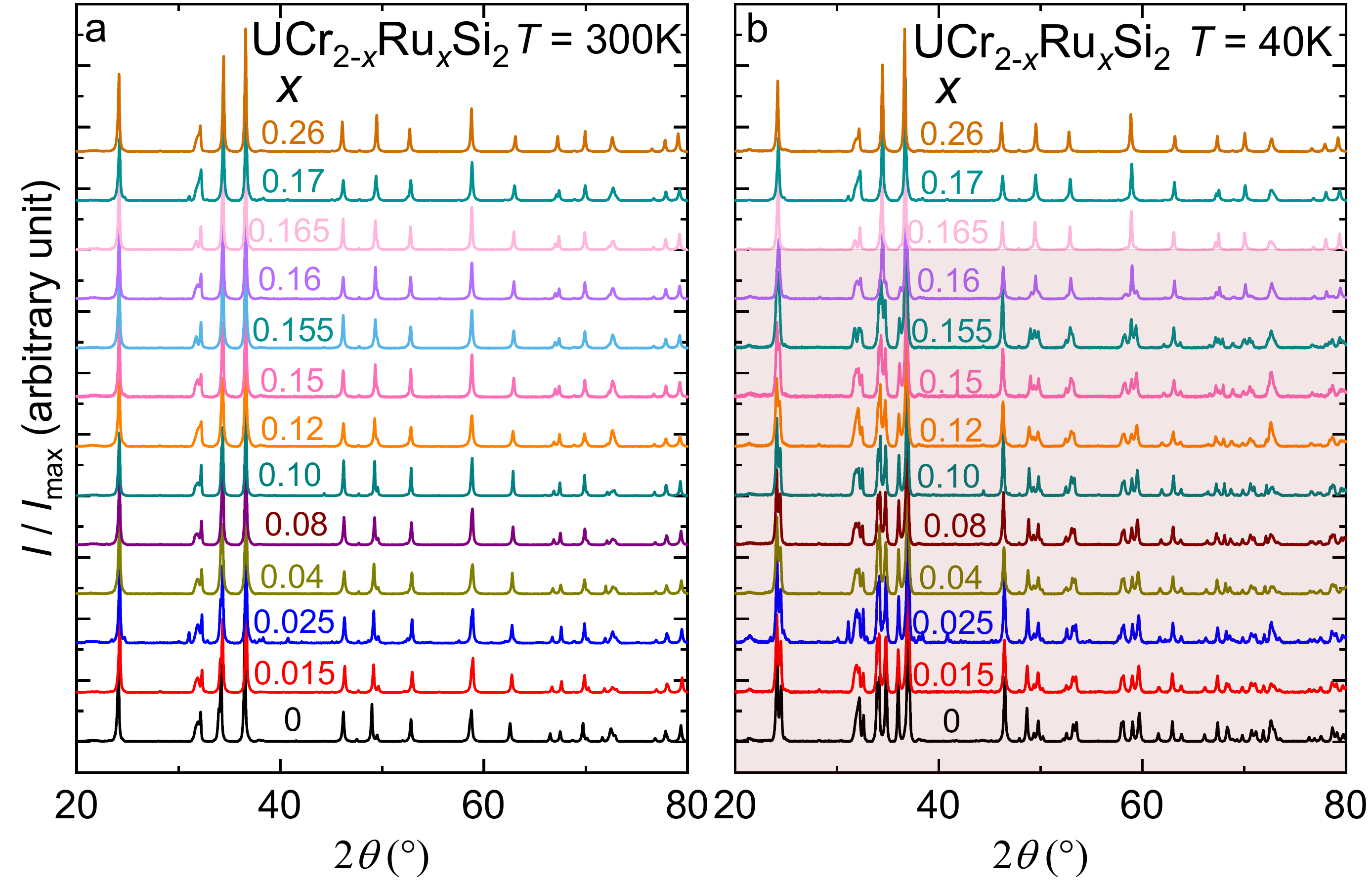}
		\caption{Powder x-ray diffraction data for UCr$_{2-x}$Ru$_x$Si$_2$ spanning concentrations 0 $<$ $x$ $<$ 0.256 and at temperatures $T$ (a) 300 K and (b) 40 K. Results at $T$ $=$ 10 K are similar to those seen at 40 K. The transition from the tetragonal to the monoclinic structure is evidenced by splitting of peaks between 300 K and 40 K, as highlighted by the shaded box in panel b.
		}
		\label{waterfall}
	\end{center}
\end{figure*}

\begin{figure*}[!tht]
	\begin{center}
		\includegraphics[width=1\linewidth]{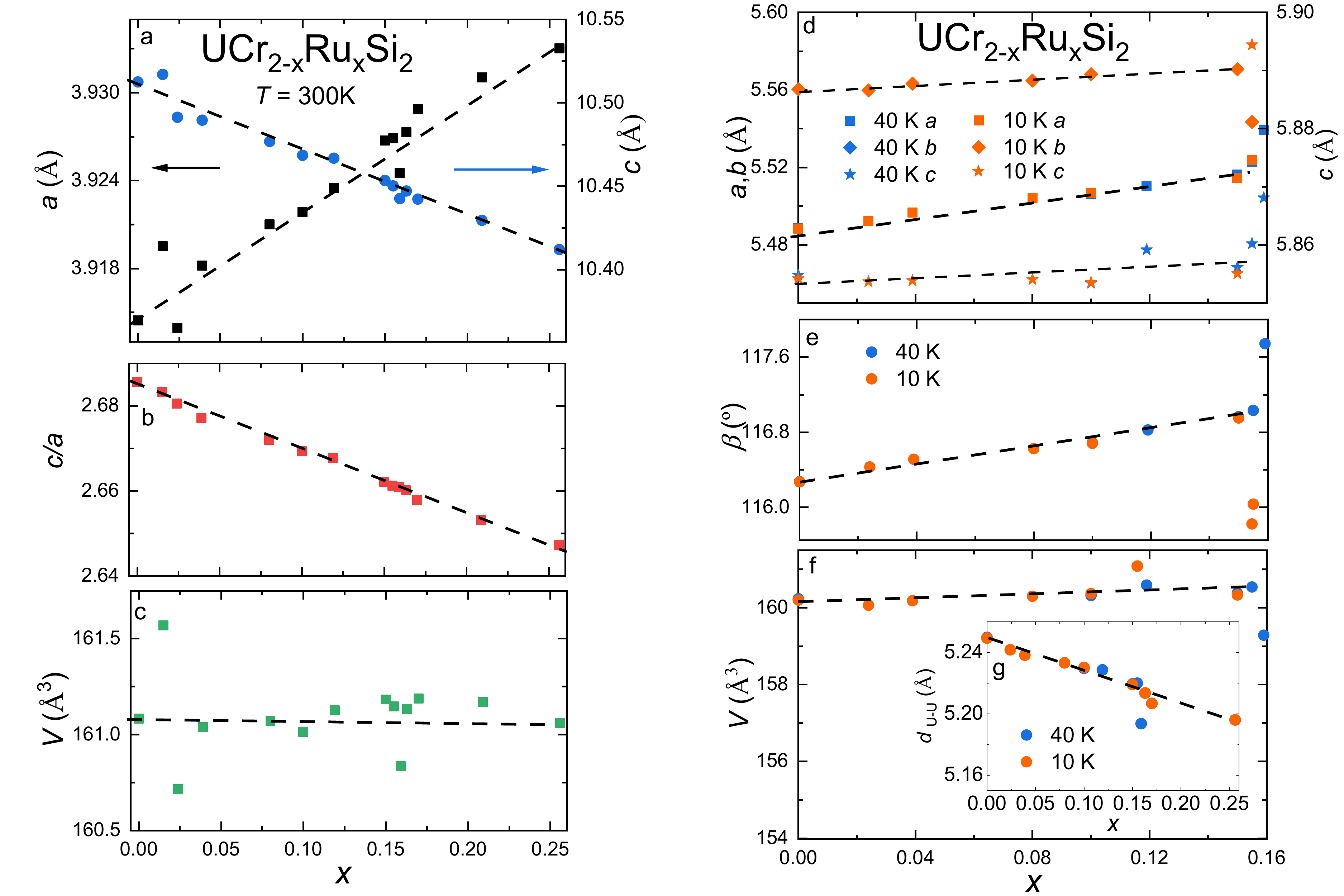}
		\caption{ Summary of structural quantities determined from analysis of powder x-ray diffraction measurements of UCr$_{2-x}$Ru$_x$Si$_2$ at temperatures of 300 K (panels a-c), 40 K, and 10 K (panels d-g). The dashed lines are guide to the eye. (a) The lattice constants, $a(x)$ (left axis) and $c(x)$ (right axis) for the 300 K tetragonal structure. (b) The $c/a$ ratio for the 300 K tetragonal structure. (c) The unit cell volume $V(x)$ for the 300 K tetragonal structure. (d) Lattice constants $a$($x$), $b$($x$), and $c$($x$) for the low temperature monoclinic structure. (e) The monoclinic distortion  angle $\beta$($x$) for the low temperature monoclinic structure. (f) The unit cell volume $V(x)$ for the low temperature monoclinic structure. (g) The interlayer U-U spacing for the monoclinic structure. Above $x$ = 0.16, the U-U spacing is considered as $c$/2 from the tetragonal structure at low temperature.
		}
		\label{xrd}
	\end{center}
\end{figure*}

To clarify the impact of Cr $\rightarrow$ Ru substitution on the structure, powder XRD measurements at $T$ $=$ 300 K,  40 K  and 10 K were performed (Fig.~\ref{waterfall} and Fig.~\ref{xrd}); i.e., spanning the $x$ $=$ 0 structural phase transition. As shown in panel a, the high temperature tetragonal ThCr$_2$Si$_2$-type structure persists across the entire substitution series up to $x$ = 0.26, without evidence for chemical phase separation or formation of impurity phases. At 300 K, fits to the data yield lattice constants $a$ and $c$, which increase and decrease with increasing $x$, respectively, while the unit cell volume $V$ remains nearly constant (Fig.~\ref{xrd}a-c). Measurements at $T$ $=$ 40 K (Fig.~\ref{waterfall}b and 10 K, not shown) next reveal that, up to $x$ = 0.16, there is a clear splitting of the peaks due to the tetragonal $\rightarrow$ monoclinic structural phase transition. For larger $x$ the splitting abruptly disappears and the ThCr$_2$Si$_2$ symmetry persists down to $T$ $=$ 10 K, suggesting that the structural phase transition collapses near $x$ = 0.16.  Within the monoclinic phase (Fig.~\ref{xrd}d-g), the lattice constants ($a$, $b$, $c$) and the distortion angle $\beta$ increase with increasing $x$. Similar to what is seen in the tetragonal phase, the spacing between the U-U atoms decreases with increasing $x$ (Fig.~\ref{xrd}g). 

\begin{figure}[!tht]
	\begin{center}
		\includegraphics[width=1\linewidth]{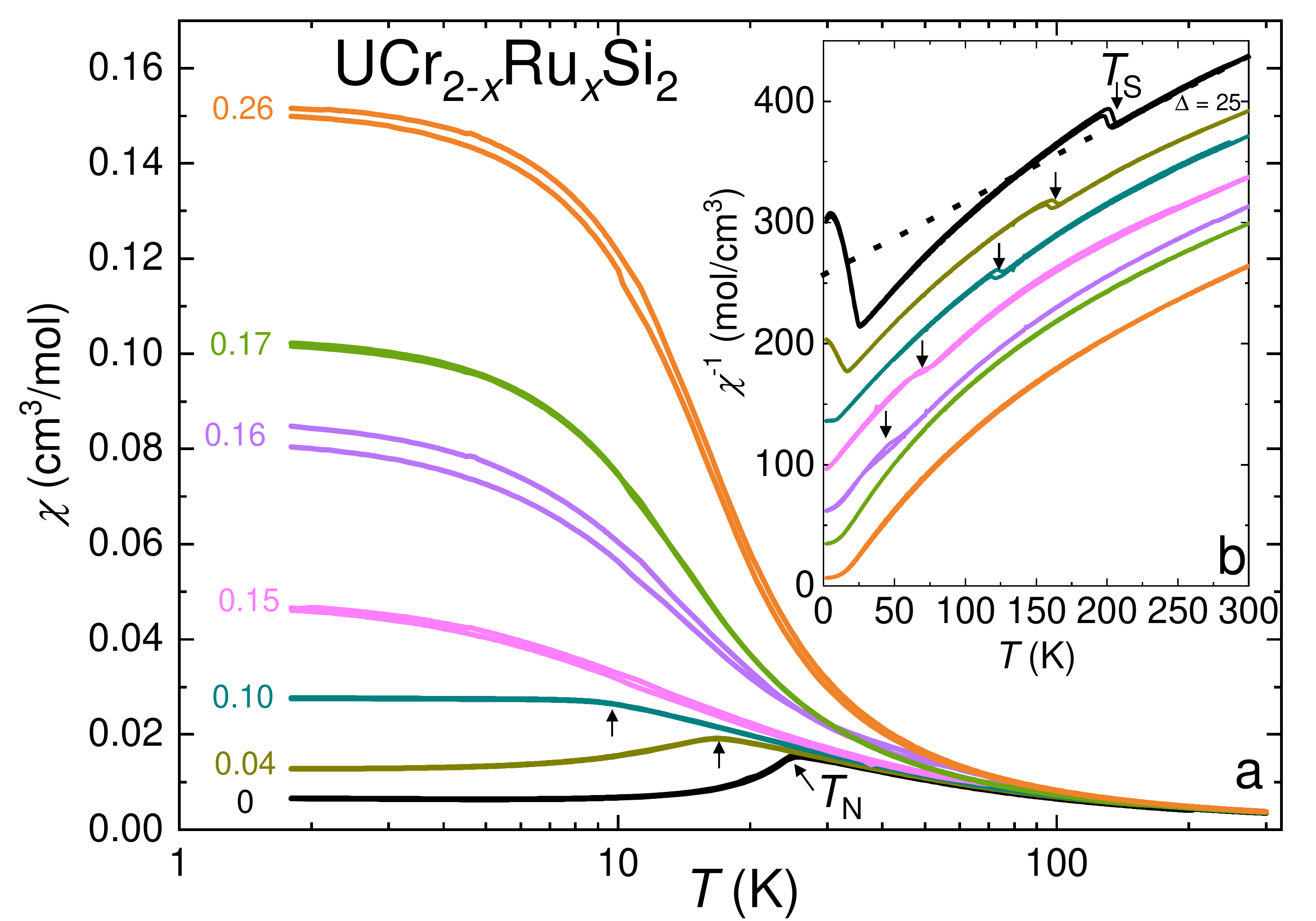}
		\caption{ (a)The magnetic susceptibility $\chi$ vs. temperature $T$ for measured in an applied magnetic field $H$ = 0.5 T for select concentrations spanning the UCr$_{2-x}$Ru$_x$Si$_2$ chemical substitution series. The antiferromagnetic ordering temperatures are indicated by arrows as $T_{\rm{N}}$. (b) The inverse magnetic susceptibility $\chi$$^{-1}$ vs. temperature $T$ for measured in an applied magnetic field $H$ = 0.5 T for select concentrations spanning the UCr$_{2-x}$Ru$_x$Si$_2$ chemical substitution series. The structural ordering temperatures is indicated by arrows as $T_{\rm{S}}$.
		}
		\label{chi}
	\end{center}
\end{figure}

The temperature dependent polycrystalline average magnetic susceptibility $\chi(T)$ data for selected concentrations are shown in Fig.~\ref{chi}. For $x$ $=$ 0, the data compare favorably to the ideal polycrystalline average $\chi_{\rm{avg}}(T)$ = (2 * $\chi_{\rm{ab}}$ + $\chi_{\rm{c}}$) /3 that is calculated from results for a single crystal specimen where $\chi_{\rm{ab}}$ and $\chi_{\rm{c}}$ are the magnetic susceptibilities when the magnetic field is applied in the $ab$-plane and along the $c$-axis, respectively (Fig. S4). The $x$ = 0 results are similar to earlier reports, where a nearly Curie-Weiss $T$-dependence is observed starting from 300 K (Fig.~\ref{chi}b). Fits to the data reveal that an effective magnetic moment $\mu_{\rm{eff}}$ $\approx$ 3.62 $\mu_{\rm{B}}$/U and the Curie-Weiss temperature $\theta$ $\approx$ -185 K. Note that the value of $\mu_{\rm{eff}}$ is similar to that seen for other intermetallics with trivalent or tetravalent uranium, but that this value is larger than what was seen in earlier reports.~\cite{Matsuda1_2002,Matsuda1_2003} The structural phase transition appears as a hysteretic feature at $T_{\rm{S}}$ $=$ 205 K, after which the CW behavior continues until the system orders antiferromagnetically near $T_{\rm{N}}$ $=$ 25 K. As shown in other measurements, both $T_{\rm{S}}$ and $T_{\rm{N}}$ are suppressed with increasing $x$. Also noteworthy is that the underlying CW temperature dependence is undisturbed by substitution, suggesting that the uranium valence state and the hybridization with the conduction electrons is unchanged. We also note that once the antiferromagnetic order is fully suppressed for $x$ $>$ 0.1, the low temperature $\chi(T)$ increases with increasing $x$, suggesting the emergence of ferromagnetic fluctuations.

\begin{figure}[!!!!!!!!!!!!!!!!!!!!!!!tht]
	\begin{center}
		\includegraphics[width=1\linewidth]{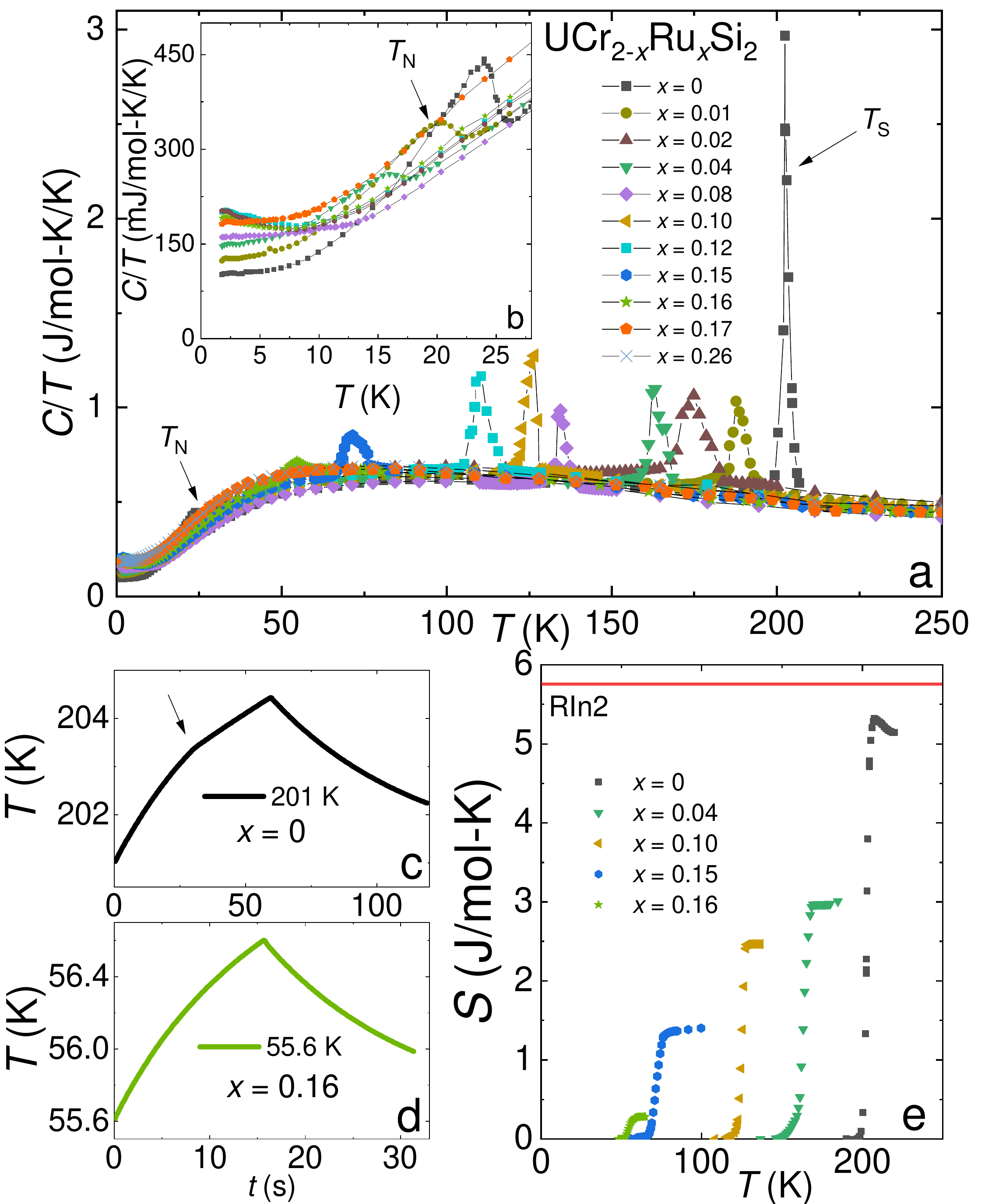}
		\caption{(a) The heat capacity divided by temperature $C$/$T$ vs $T$ for UCr$_{2-x}$Ru$_x$Si$_2$ emphasizing the structural phase transition at $T_{\rm{S}}$. (b) Low temperature $C$/$T$ for selected concentrations showing both the antiferromagnetic phase transition at $T_{\rm{N}}$ and the low temperature upturn that are described in the text. (c) and (d) relaxation curves around the structural phase transition temperature for $x$ = 0 and $x$ = 0.16. In panel c the arrow indicates the kink in the heating curve due to the latent heat of the first order phase transition. A similar feature is not detected for $x$ = 0.16. (e) The entropy associated with the structural phase transition for selected $x$, which was determined as described in the text.   
		}
		\label{HC}
	\end{center}
\end{figure}

The temperature dependence of the heat capacity divided by temperature $C/T$ further exposes the structural phase transition and its impact on the electronic and magnetic properties (Fig.~\ref{HC}). At $x$ $=$ 0 it appears as a sharp first-order peak near $T_{\rm{S}}$ $=$ 205 K and with increasing $x$ it is monotonically suppressed down to 55 K at $x$ $=$ 0.16, after which it abruptly disappears before $x$ = 0.165. A close inspection of the thermal relaxation curves reveals that for $x$ = 0 there is a kink in the temperature rising curve, consistent with there being a latent heat. This feature weakens with increasing $x$, and is not easily seen for $x$ = 0.16 (Fig.~\ref{HC}c and d). This implies that the transition becomes less strongly first order as $T_{\rm{S}}$ is suppressed, possibly as a result of disorder effects. As shown in Fig.~\ref{HC}e, the entropy recovered at $T_{\rm{S}}$ also grows smaller with increasing $x$, and is significantly reduced as $x$ approaches 0.16. 

Cr $\rightarrow$ Ru substitution also strongly modifies the signature of the antiferromagnetic ordering in $C/T$, which is seen as a second order lambda like peak at $T_{\rm{N}}$ = 25 K for $x$ $=$ 0 (Fig.~\ref{HC}b).~\cite{Matsuda1_2002} In particular, $T_{\rm{N}}$ moves to lower temperatures until it disappears near $x$ $=$ 0.08. Over this $x$-range, its shape is preserved but its overall size grows smaller, indicating that much of the associated magnetic entropy is lost or moves below $T_{\rm{N}}$. Once the antiferromagnetism is no longer observed, the low temperature $C$/$T$ exhibits a weak increase with decreasing $T$ which becomes maximal near $x$ $\approx$ 0.15. For larger $x$ the divergence weakens and finally tends to saturate at low temperature as expected for a Fermi liquid. Note that although $\chi(T)$ provides evidence for ferromagnetic fluctuations over this $x$ - range, the heat capacity data do not reveal any long range ordering: i.e., there are no sharp or lambda-like features that would indicate first or second order phase transitions, respectively.

\section{\label{sec:level1}Discussion}

\begin{figure}[!tht]
	\begin{center}
		\includegraphics[width=1\linewidth]{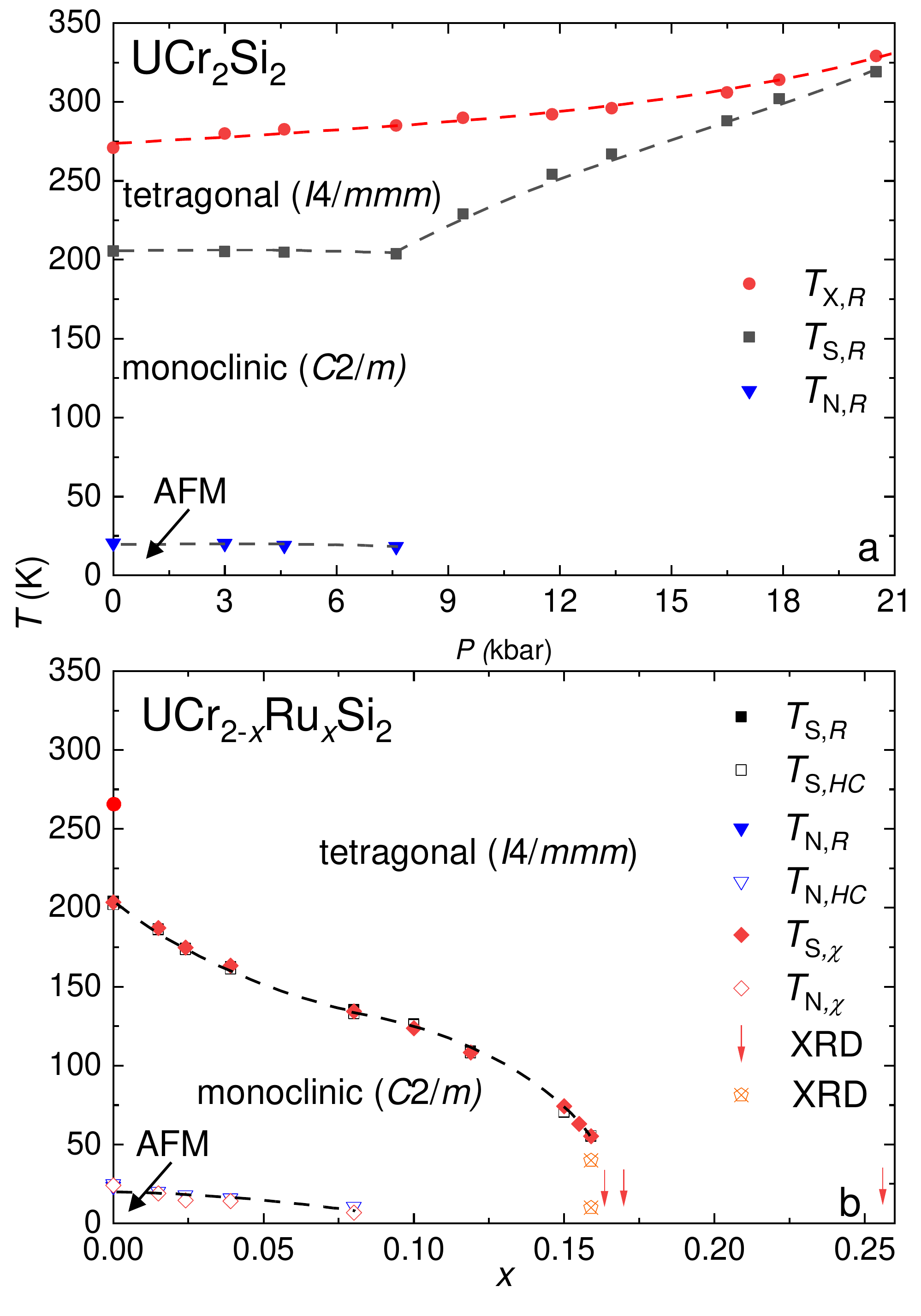}
		\caption{(a) Temperature $T$ vs pressure $P$ phase diagram for UCr$_2$Si$_2$ constructed from electrical resistance $R$ measurements under applied pressure. The unidentified phase transition at $T_{\rm{X}}$, the structural phase transition at $T_{\rm{S}}$, and the antiferromagnetic phase transition at $T_{\rm{N}}$ are shown. (b) Temperature $T$ vs. concentration $x$ phase diagram for UCr$_{2-x}$Ru$_x$Si$_2$ for $x$ $=$ 0 $-$ 0.26 constructed from electrical resistance $R$, powder x-ray diffraction, magnetic susceptibility $\chi$, and heat capacity $C$ measurements. The arrows represent results from XRD measurements where the tetragonal phase persists down to $T$ $=$ 10 K. At these same concentrations, $R$ and $C$ measurements show that there are no phase transitions for $T$ $>$ 500 mK. 
		}
		\label{phase}
	\end{center}
\end{figure}

Fig.~\ref{phase} compares the $T-P$ and $T-x$ phase diagrams compiled from the above measurements, where hydrostatic pressure and non-isoelectronic chemical substitution have profoundly different impacts on the structural and magnetic phase transitions. In the former case, $T_{\rm{X}}$, $T_{\rm{S}}$, and $T_{\rm{N}}$ are nearly constant up to pressures near 7.6 kbar, showing that small perturbations that isotropically compress the unit cell volume minimally impact the system. The data then reveal an evolution in the structural and magnetic phases for $P$ $>$ 7.6 kbar. While further measurements such as X-Ray diffraction are needed to fully understand the phase diagram, we point out that this behavior might be expected given that hydrostatic pressure should evenly decrease both the $ab$ and $c$ directions, which roughly mimics the influence of decreasing temperature.~\cite{Lemoine_2018} As a result, the energy scale of the structural phase transition would only be expected to increase with $P$. 

In contrast, Cr $\rightarrow$ Ru substitution immediately destroys the feature at $T_{\rm{X}}$, while $T_{\rm{S}}$ and $T_{\rm{N}}$ are rapidly suppressed and collapse towards zero temperature near $x_{\rm{c, N}}$ $=$ 0.08 and $x_{\rm{c, S}}$ $=$0.16, respectively. As reported earlier, the strong Cr-Si and Si-Si bonds directly contribute to the lattice rigidity, and therefore to the relative stabilities of the high-temperature tetragonal and low-temperature monoclinic phases.~\cite{Lemoine_2018} When Ru is introduced, the Cr(Ru)-Si bond is stretched due to the size difference between Cr and Ru, leading to an expansion in the $a-b$ plane and a compression along the $c$-axis which anisotropically opens the voids where the U atoms are located. This apparently destabilizes the monoclinic phase (driving it to lower temperatures) and increases the $\beta$ angle. What is unclear in this process is the role of the bonding between the uranium and its surrounding environment. To clarify this, we point out (i) that earlier work shows that U is loosely bonded to the atoms surrounding it~\cite{Lemoine_2018}, (ii) that $\chi(T)$ measurements show that the uranium $f$-electron valence does not noticeably change between the two structural phases, which would be expected if a change of the $f$-state itself were responsible for the lattice instability and (iii) that the underlying Kondo lattice behavior appears unchanged both with applied pressure and under chemical substitution, showing that the hybridization between the $f$- and conduction electrons is robust in the $T-P-x$ phase space. Therefore, it is likely that the structural phase transition is driven by the characteristics of the strong Cr-Si and Si-Si bonds, which are anisotropically tuned by the introduction of Ru.

Structural tuning subsequently has an important impact on the low temperature electronic/magnetic state. First, both phase diagrams show that the original antiferromagnetic order is only preserved when the monoclinic phase is unambiguously present. In particular, (i) it becomes unidentifiable in $\rho(T)$ as $T_{\rm{S}}$ begins to increase above 7.6 kbar and (ii) it is suppressed as $T_{\rm{S}}$ decreases and eventually is replaced by strengthening ferromagnetic fluctuations that span $x_{\rm{c, S}}$ $=$ 0.16. The collapse of $T_{\rm{S}}$ towards zero temperature at $x_{\rm{c, S}}$ is also of interest because it raises the possibility of there being a structural quantum critical point in this region. However, several features show that $T_{\rm{S}}$ remains first order as it is suppressed including that (i) the approach of the phase boundary towards zero temperature is nearly vertical and (ii) hysteresis is observed up to $x_{\rm{c, S}}$ $=$ 0.16. This prevents the occurrence of diverging quantum critical fluctuations, but might offer insights about how to proceed in future work.

Taken together, these results provide a window into the physics of an unusual quantum material: a Kondo lattice with tunable structural and magnetic instabilities.~\cite{Pfleiderer09,Maple10,Andy_2016,Greve_2010,Handunkanda_2015,Goh_2015,Goh_15} It remains important to fully establish the mechanism for the structural phase transition, where calculations and measurement of the phonon band structure and/or measurements such as X-Ray absorption Spectroscopy or ARPES will be useful. Also important is to investigate the feature at $T_{\rm{X}}$, which may be related to charge density wave features seen in $R$Pt$_2$Si$_2$($R$ = Y, La, Nd, Lu)~\cite{LaPt2Si2, RPt2Si2} and UPt$_2$Si$_2$~\cite{UPt2Si2}. Finally, the $T-P$ and $T-x$ phase diagrams provide guidance for further efforts to induce a magnetic or structural quantum critical point. For instance, measurements under uniaxial pressure would be useful to determine whether the decrease in the $c/a$ ratio is the dominant tuning parameter in the Cr $\rightarrow$ Ru substitution series, while chemical substitution that uniformly expands the lattice (e.g., Cr $\rightarrow$ Mo or W) would be expected to suppress $T_{\rm{S}}$. This might eventually provide access to phenomena that are distinct from what is seen in other systems with structural quantum phase transitions: e.g., such as LaCu$_{6-x}$Au$_x$, ScF$_3$, and (Ca$_x$Sr$_{1-x}$)$_3$Rh$_4$Sn$_{13}$, where the structural instability arises solely from the freezing of a soft phonon mode,~\cite{Andy_2016,Greve_2010,Handunkanda_2015,Goh_2015,Goh_15,Klintberg_12,Biswas_15} or the Fe-based superconductors, where strong electronic correlations originate from the Fe $d$-electron states and the magnetic/structural order are closely tied together.~\cite{Cruz_2008,Chu_2010,Fernandes_2014}

\section{\label{sec:level1}Conclusions}
We have studied the influence of applied pressure and Cr $\rightarrow$ Ru chemical substitution in UCr$_2$Si$_2$, where the former semi-uniformly reduces the unit cell and the latter results in a decrease along the $c$-axis and an expansion in the $ab$-plane. While hydrostatic compression increases the structural ordering temperature, disrupts the antiferromagnetism, and may induce a different structural order at high pressures, Cr $\rightarrow$ Ru substitution suppresses $T_{\rm{N}}$ towards zero temperature near $x_{\rm{c, N}}$ $\approx$ 0.08 and suppresses the structural phase transition $T_{\rm{S}}$ until it disappears near $x_{\rm{c, S}}$ $\approx$ 0.16, after which the tetragonal phase persists to low temperatures. In the region where $T_{\rm{S}}$ is maximally suppressed it remains a first order phase transition, which prevents the occurrence of a quantum critical fluctuations of the order parameter. This study provides insights into the structural phase transition and magnetism of UCr$_2$Si$_2$, uncovers a high temperature phase at $T_{\rm{X}}$, and makes progress towards uncovering an electronic/structural quantum phase transition in an $f$-electron system, where alternative tuning strategies are suggested.

\section{\label{sec:level1}Acknowledgements}
A portion of this work was performed at the National High Magnetic Field Laboratory (NHMFL), which is supported by National Science Foundation Cooperative Agreement No. DMR-1157490 and DMR-1644779, and the State of Florida. Research of RB, YL, DG, were supported in part by the Center for Actinide Science and Technology, an Energy Frontier Research Center funded by the U.S. Department of Energy (DOE), Office of Science, Basic Energy Sciences (BES), under Award Number DE-SC0016568. PJWM and JD acknowledge funding by the European Research Council (ERC) under the European Union’s Horizon 2020 research and innovation program (grant no. 715730, MiTopMat).

\clearpage

\begin{center}

\section{supplementary materials}

\begin{figure*}[!tht]
	\begin{center}
		\includegraphics[width=1\linewidth]{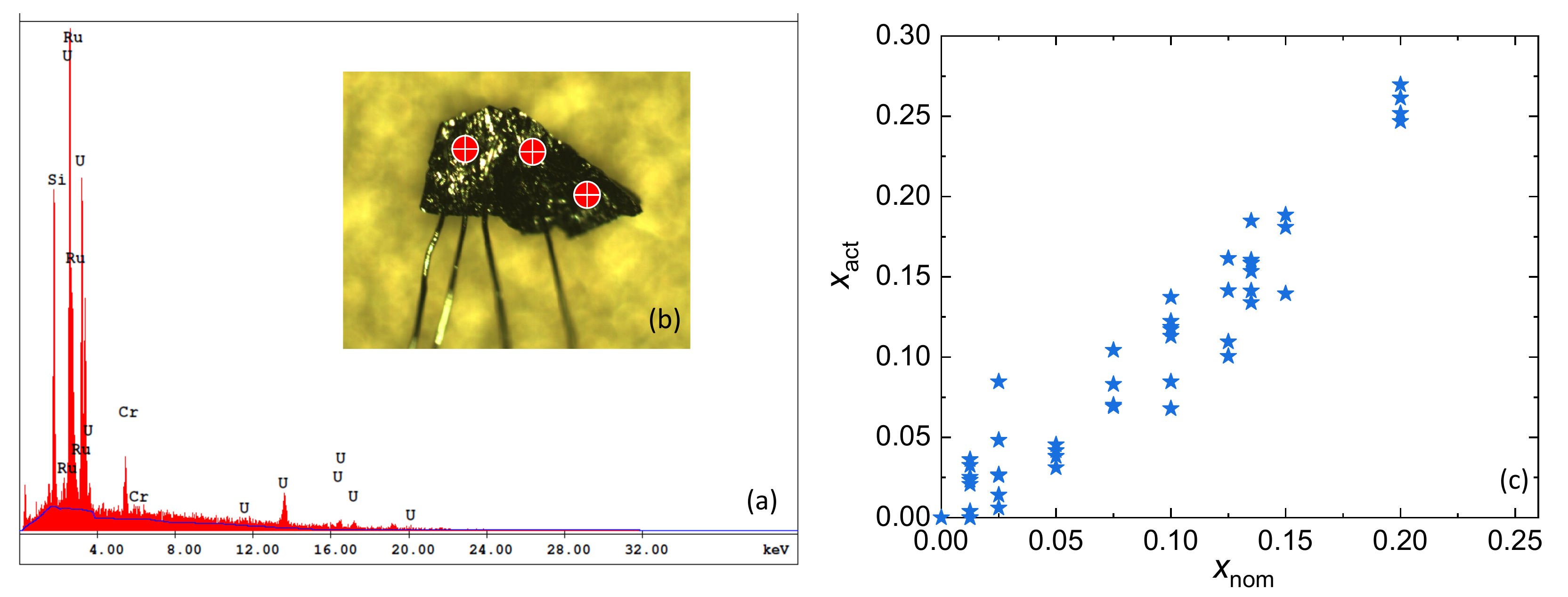}
		\caption{\textbf{Fig. S1. Summary of energy dispersive spectroscopy data for chemical analysis for the UCr2-xRuxSi2 series:} (a) An example of energy dispersive spectroscopy (EDS) data set for a UCr$_{2-x}$Ru$_x$Si$_2$ specimen. (b) Samples were attached to conducting carbon tape and measurements were made at several spots, as shown by the red circles. (c) Plot of the actual concentration $x_{\rm{act}}$ vs the nominal concentration $x_{\rm{nom}}$. Within the manuscript, $x_{\rm{act}}$ is defined as the average value for the multiple measurements that were performed on a given sample.
		}
		\label{waterfall}
	\end{center}
\end{figure*}

\begin{figure*}[!tht]
	\begin{center}
		\includegraphics[width=0.7\linewidth]{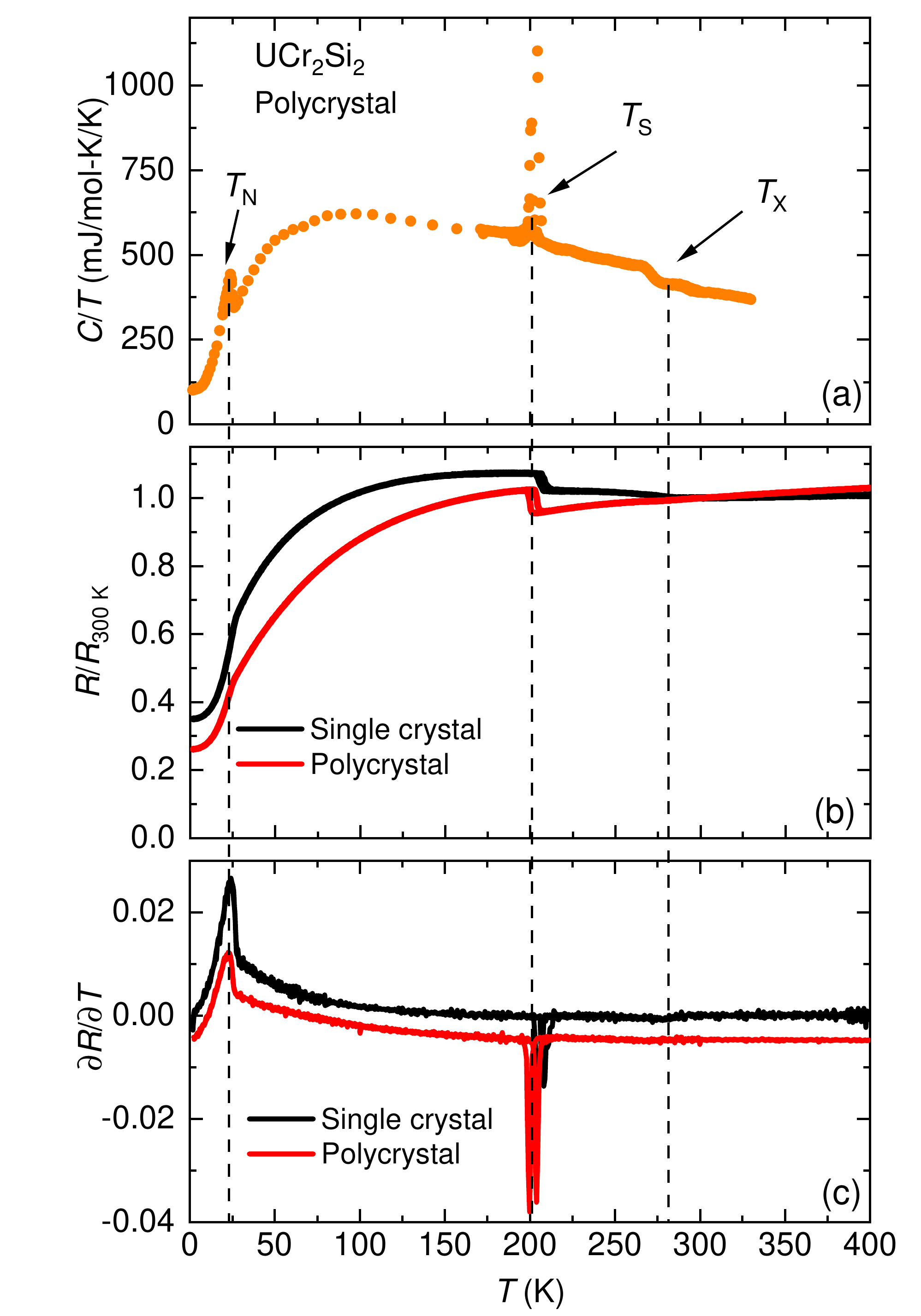}
		\caption{\textbf{Fig. S2. Summary of heat capacity and electrical transport measurements for UCr$_2$Si$_2$:} (a) Temperature dependent heat capacity measurements for polycrystalline UCr$_2$Si$_2$, where the newly identified ordering temperature ($T_{\rm{X}}$), the structural phase transition ($T_{\rm{S}}$) and antiferromagnetic transition ($T_{\rm{N}}$) are indicated. (b) $R/R_{300 K}$ vs $T$ for poly- and single crystal specimens of UCr$_2$Si$_2$. For the single crystal, the electrical current is applied in the $ab$ plane. (c) First derivative of $R$ with respect to $T$ for poly- and single crystal specimens. 
		}
		\label{S2}
	\end{center}
\end{figure*}

\begin{figure*}[!tht]
	\begin{center}
		\includegraphics[width=1\linewidth]{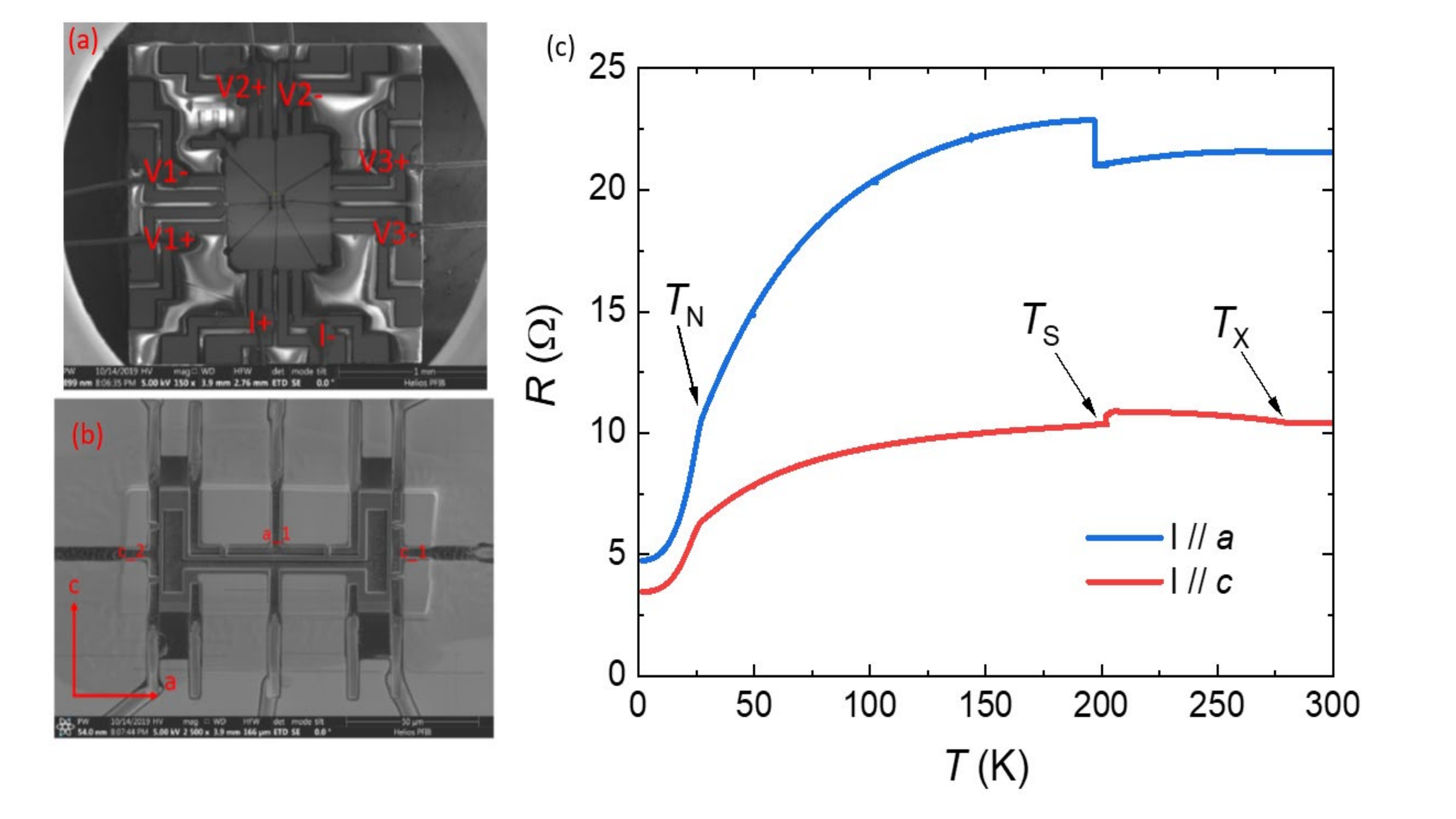}
		\caption{\textbf{Fig. S3. Summary of temperature dependent electrical transport data for single crystal of UCr$_2$Si$_2$ that was microstructured using the focused ion beam milling method:} (a) A view of the full microstructure as prepared for electrical resistivity measurements. The electrical current $I$ and voltage $V$ contacts are labeled. (b) Zoom view of the microstructure, where a and c are the crystallographic axes. (c) Resistance $R$ vs temperature $T$ measurement of the microstructure, where the $I$ is driven through the $c$ and $a$ axes of the crystal. The newly identified ordering temperature ($T_{\rm{X}}$), structural phase transition ($T_{\rm{S}}$) and antiferromagnetic transition ($T_{\rm{N}}$) are indicated.
		}
		\label{S3}
	\end{center}
\end{figure*}

\begin{figure*}[!tht]
	\begin{center}
		\includegraphics[width=0.7\linewidth]{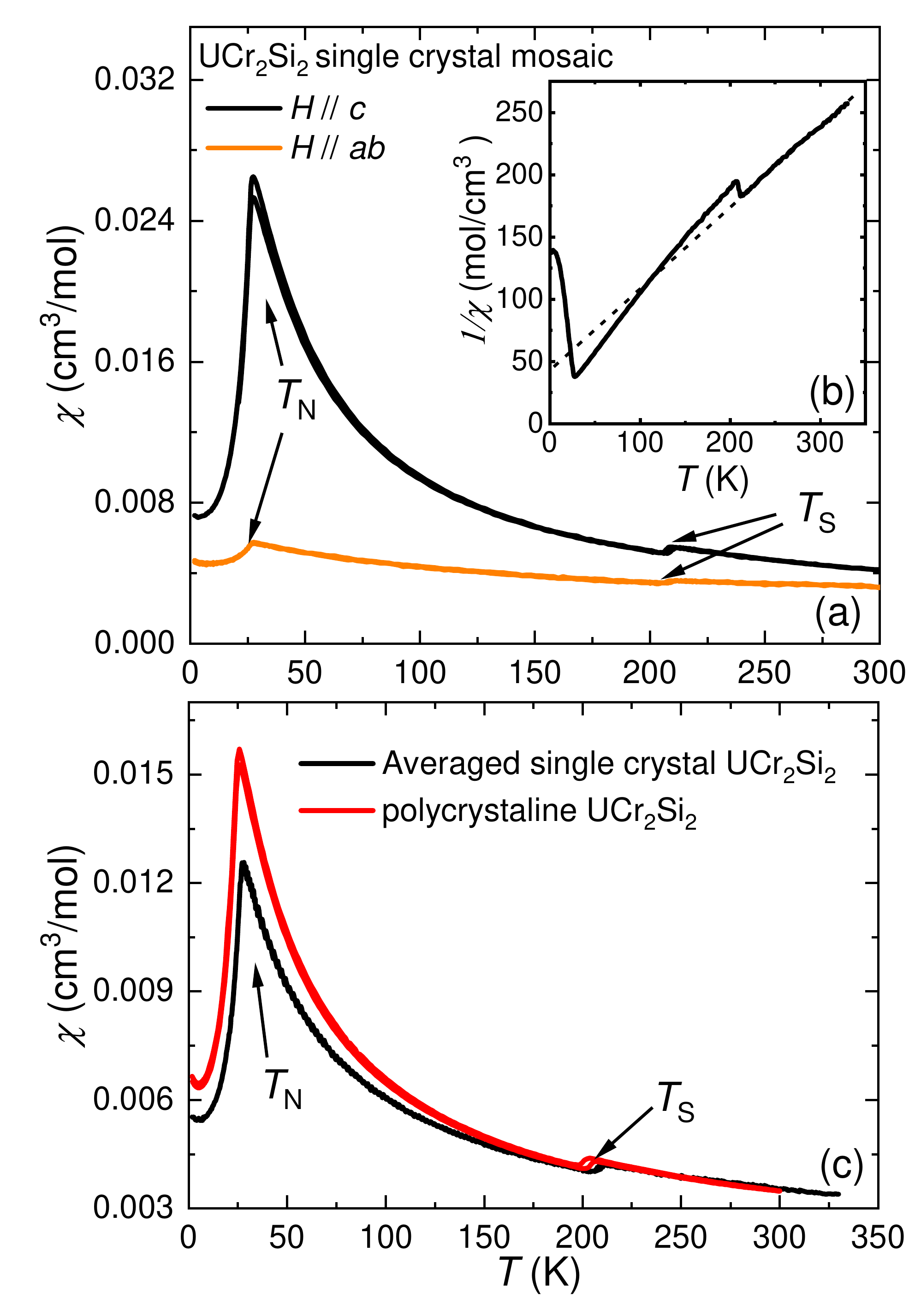}
		\caption{\textbf{Fig. S4. Temperature dependent DC magnetic susceptibility for UCr2Si2:} (a) Magnetic susceptibility $\chi(T)$= $M/H$ collected for a single crystal mosaic with $H$ = 1 tesla applied parallel (//) to the ab and c crystalline axes. (b) Inverse magnetic susceptibility $\chi$$^{-1}$$(T)$ for $H$ // $c$ where the dotted line is a Curie-Weiss fit to the data above $T_{\rm{S}}$. (c) Comparison between $\chi(T)$ for a polycrystalline specimen and the ideal polycrystalline average $\chi_{\rm{avg}}$ calculated from results for a single crystal specimen. $\chi_{\rm{avg}}$ = (2* $\chi_{\rm{ab}}$ + $\chi_{\rm{c}}$)/3, where $\chi_{\rm{ab}}$ and $\chi_{\rm{c}}$ are the magnetic susceptibilities when the magnetic field is applied in the ab-plane and along the $c$-axis, respectively.
		}
		\label{S4}
	\end{center}
\end{figure*}

\end{center}

\end{document}